\documentclass{aa} 
\usepackage{graphicx}
\usepackage{natbib}
\bibpunct{(}{)}{;}{a}{}{,}
\usepackage{verbatim}
\usepackage{txfonts}
\usepackage[T1]{fontenc}
\newcommand{\ls}[3][]{\ensuremath{^{#2}\mathrm{#3}_{#1}}}
\newcommand{\ie}{i.e.\ }
\begin{document}
\title{The FERRUM Project: Experimental and theoretical transition rates of forbidden [\ion{Sc}{ii}] lines and radiative lifetimes of metastable \ion{Sc}{ii} levels \thanks{Based on observations made with the NASA/ESA Hubble Space Telescope, obtained at the Space Telescope Science Institute. STScI is operated by the Association of Universities for Research in Astronomy, Inc. under NASA contract NAS 5-26555. Laboratory data is obtained at the CRYRING facility at the Manne Siegbahn Laboratory (MSL), Stockholm University, Sweden.}}
\author{H.~Hartman\inst{1} \and J.~Gurell\inst{2} \and P.~Lundin\inst{2} \and P.~Schef\inst{2} \and A.~Hibbert\inst{3} 
\and H.~Lundberg\inst{4} \and S.~Mannervik\inst{2} \and L.-O.~Norlin\inst{5} \and P.~Royen\inst{2}}
\titlerunning{$A$-values for forbidden \ion{Sc}{ii} lines and metastable lifetimes}
\offprints{Henrik Hartman, \email{Henrik.Hartman@astro.lu.se}}
\institute{Lund Observatory, Lund University, Box 43, SE-221\,00 Lund, Sweden 
\and Department of Physics, Stockholm University, AlbaNova University Center, SE-106\,91 Sweden 
\and Depertment of Applied Mathematics and Theoretical Physics, Queen's University, Belfast BT7 1NN, Northern Ireland
\and Department of Physics, Lund Institute of Technology, Box 118, SE-221\,00 Lund, Sweden
\and Department of Physics, Royal Institute of Technology, AlbaNova University Center, SE-106\,91 Stockholm, Sweden}
\date{Received <date> / Accepted <date>}

\abstract{In many plasmas, long-lived metastable atomic levels are depopulated by collisions (quenched) before they decay radiatively. In low-density regions, however, the low collision rate may allow depopulation by electric dipole (E1) forbidden radiative transitions, so-called forbidden lines (mainly M1 and E2 transitions) . If the atomic transition data is known, these lines are indicators of physical plasma conditions and used for abundance determination.}
{Transition rates can be derived by combining relative intensities between the decay channels, so-called branching fractions (BFs), and the radiative lifetime of the common upper level. We use this approach for forbidden [\ion{Sc}{ii}] lines, along with new calculations.}
{Neither BFs for forbidden lines, nor lifetimes of metastable levels, are easily measured in a laboratory. Therefore, astrophysical BFs measured in Space Telescope Imaging Spectrograph (STIS) spectra of the strontium filament of \object{Eta Carinae} are combined with lifetime measurements using a laser probing technique on a stored ion-beam (CRYRING facility, MSL, Stockholm). These quantities are used to derive the absolute transition rates ($A$-values). New theoretical transition rates and lifetimes are calulated using the CIV3 code.}
{We report experimental lifetimes of the \ion{Sc}{ii} levels 3d$^2$~a$^3$P$_{0,1,2}$ with lifetimes 1.28, 1.42, and 1.24~s, respectively, and transition rates for lines from these levels down to 3d4s~a$^3$D in the region 8270-8390\AA. These are the most important forbidden [ScII] transitions. New calculations for lines and metastable lifetimes are also presented, and are in good agreement with the experimental data.}
{} 

\maketitle
\newpage
\section{Introduction}
Physical conditions and elemental abundances of stellar and nebular objects are derived from quantitative analysis of spectra. For \ion{H}{II} regions and gaseous nebulae, emission lines are used to derive these properties based on simple line ratios or more detailed modeling. In either case, accurate atomic data is needed to obtain reliable determinations. One important property is the radiative transition rate for spontaneous decay. 

For complex spectra, which have many lines, not all transition rates can be measured. The majority of the transition data need to be determined through calculations. Experimental data with reliable uncertainties are measured for important transitions and used to compare with the calculations.

The FERRUM project was initiated to obtain experimental data and evaluate theoretical data for the iron group elements used in astrophysical applications \citep{JDD02}. This is an international collaboration with the aim to produce reliable oscillator strengths and transition rates for transitions used in astrophysical analysis. One branch within this collaboration is to measure metastable lifetimes and, in combination with astrophysical branching fractions, derive absolute transition rates for forbidden lines. Forbidden lines are, by definition, lines that do not occur by electric dipole radiation (E1), but rather with higher multipole radiation, in particular magnetic dipole (M1) or electric quadropole (E2). Parity forbidden lines are transitions between levels of equal parity, and occur between the lowest levels, which cannot decay via E1 transitions. Forbidden lines with even higher multipole radiation can have transitions between levels of opposite parity. In this case, it is often the change in $J$-value that makes the E1 transition forbidden. Levels that can decay only by forbidden transitions are called metastable levels, and often have radiative lifetimes of the order of seconds, compared to nanoseconds for normal levels.

\begin{figure}
\includegraphics[width=12cm]{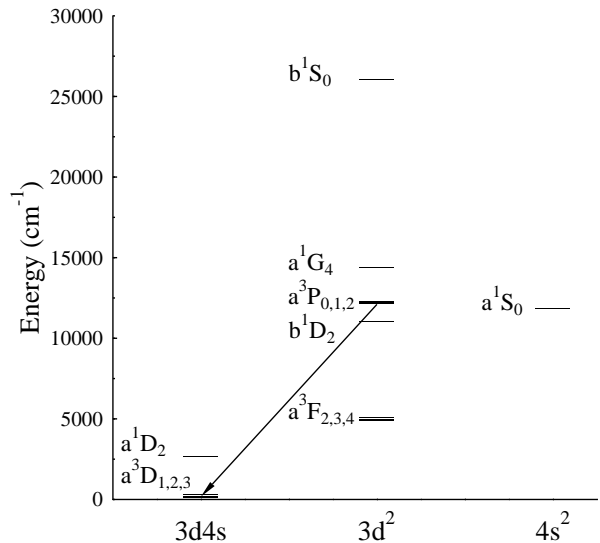} 
\caption{Partial energy level diagram of \ion{Sc}{ii}, showing the three lowest configurations. The lines measured in the present paper are 3d4s\,a\ls{3}{D}~-~3d$^2$\,a\ls{3}{P}.\label{fig:elevels} }
\end{figure}

Due to the long lifetimes of metastable levels, collisions are often the dominating deexcitation process in dense plasmas. In low-density plasmas, as in gaseous nebulae and HII regions, the radiative decay can be of similar strength or the main process. This sensitivity to density makes the forbidden lines good for density diagnostics, and also as temperature probes by simple line ratio measurements. The same collision effect occurs in the laboratory, and a low pressure environment is required to measure the lifetimes of metastable levels. In the present work, we used a storage ring with ultra-high vacuum where the ions can be stored for a long enough time to allow for measurement. Another consequence of the sensitivity to collisions is the problem of producing forbidden lines to measure branching fractions, \ie the fraction of decay in the different decay channels from a specific upper level. We use $HST$/STIS spectra of Eta Carinae to measure the lines.

Eta Carinae is one of the most massive stars known in the Galaxy. It has a violent history and has experienced several nonterminal explosions of which the so-called Great Eruption in the 1840's produced its today well-known bipolar lobes and the intervening disklike structure \citep[e.g.][]{DH97,SMG04}. One region of the disk shows a remarkable spectrum, among others forbidden [\ion{Sr}{ii}] along with many permitted and forbidden lines of iron group elements, specifically \ion{Ti}{ii} \citep{HGJ04}. This region was given the name 'Strontium Filament' (SrF) from the identification of [\ion{Sr}{ii}], which previously was not observed in any other astronomical object \citep{ZGH01}. In a previous paper \citet{HSL05} derived $A$-values for [\ion{Ti}{ii}] with these spectra. Photo-ionization modelling indicate peculiar abundance patterns and photo-excitation in this region \citep{BGI02,BHG06}.

In the spectrum of the SrF of Eta Carinae \citep{HGJ04} forbidden lines of [\ion{Sc}{ii}] were observed, \mbox{3d4s\,a$^3$D-3d$^2$a$^3$P} (Multiplet 3F of \citet{M45}). We measured the branching fractions for these lines to determine the transition rates.
In Fig.~\ref{fig:elevels}, a partial energy level diagram of \ion{Sc}{ii} is shown, based on the analysis by \citet{JL80}. Included are levels from the three lowest configurations: 3d4s, 3d$^2$, and 4s$^2$. Lines between these levels are parity forbidden. 

In Sec.~\ref{exp} we describe the experimental technique: the lifetime measurements and the derivation of astrophysical branching fractions. In Sec.~\ref{theory} we present new calculations using the CIV3 code and in Sec.~\ref{results} the results are discussed, along with uncertainties.

\section{Experimental technique} \label{exp}
A widely-used technique to derive absolute transition rates ($A$), or oscillator strengths, is the combination of the lifetime ($\tau$) of an upper level and the branching fractions (BFs) of the different decay channels from the same level:

\begin{equation} \label{eqn:A}
A_{ul} = \frac{BF_{ul}}{\tau_u}
\end{equation}
where
\begin{equation}\label{eqn:BF}
BF_{ul} = \frac{A_{ul}}{\sum_l A_{ul}} 
\end{equation}
and
\begin{equation}\label{eqn:tau}
\tau_u=\frac{1}{\sum_l A_{ul}}.
\end{equation}
The subscripts $u$ and $l$ represent the upper and lower level, respectively. 
The technique is primarily used for electric dipole transitions (E1). In that case, the lifetime is often measured with a laser induced fluorescence (LIF) technique, and the BFs are measured in a laboratory light source. For forbidden lines, this technique cannot be used since the timescale is totally different: seconds, instead of nanoseconds. 

\begin{figure} 
\resizebox{\columnwidth}{!}{\includegraphics{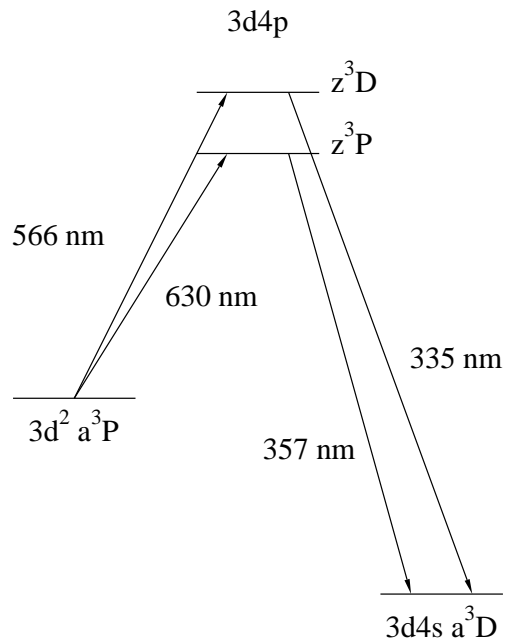}} 
\caption{Energy level diagram showing the pumping channels and major fluorescence decay used in the present experiment, with their approximate wavelengths. The laser is tuned to match the transitions to z$^3$D or z$^3$P, and the subsequent decay to a$^3$D is observed.
The level energies are not to scale.\label{fig:pump}}
\end{figure}
   
\subsection{Measurement of metastable lifetimes in \ion{Sc}{ii}}
At the CRYRING facility, a technique has been developed to measure lifetimes of metastable levels by laser probing of a stored ion beam. The ultra-high vacuum ($\sim$10$^{-11}$ mbar) makes collisons rare, and the metastable ions survive for a time comparable to several lifetimes. The laser probing technique (LPT) and its later developments have been decribed in a number of publications \citep{LAN99,M03,MEL05}. Here we only discuss the general idea of the technique and issues specific to the present measurement, and refer to previous publications for a more detailed description.

\begin{figure} 
\resizebox{\columnwidth}{!}{\includegraphics{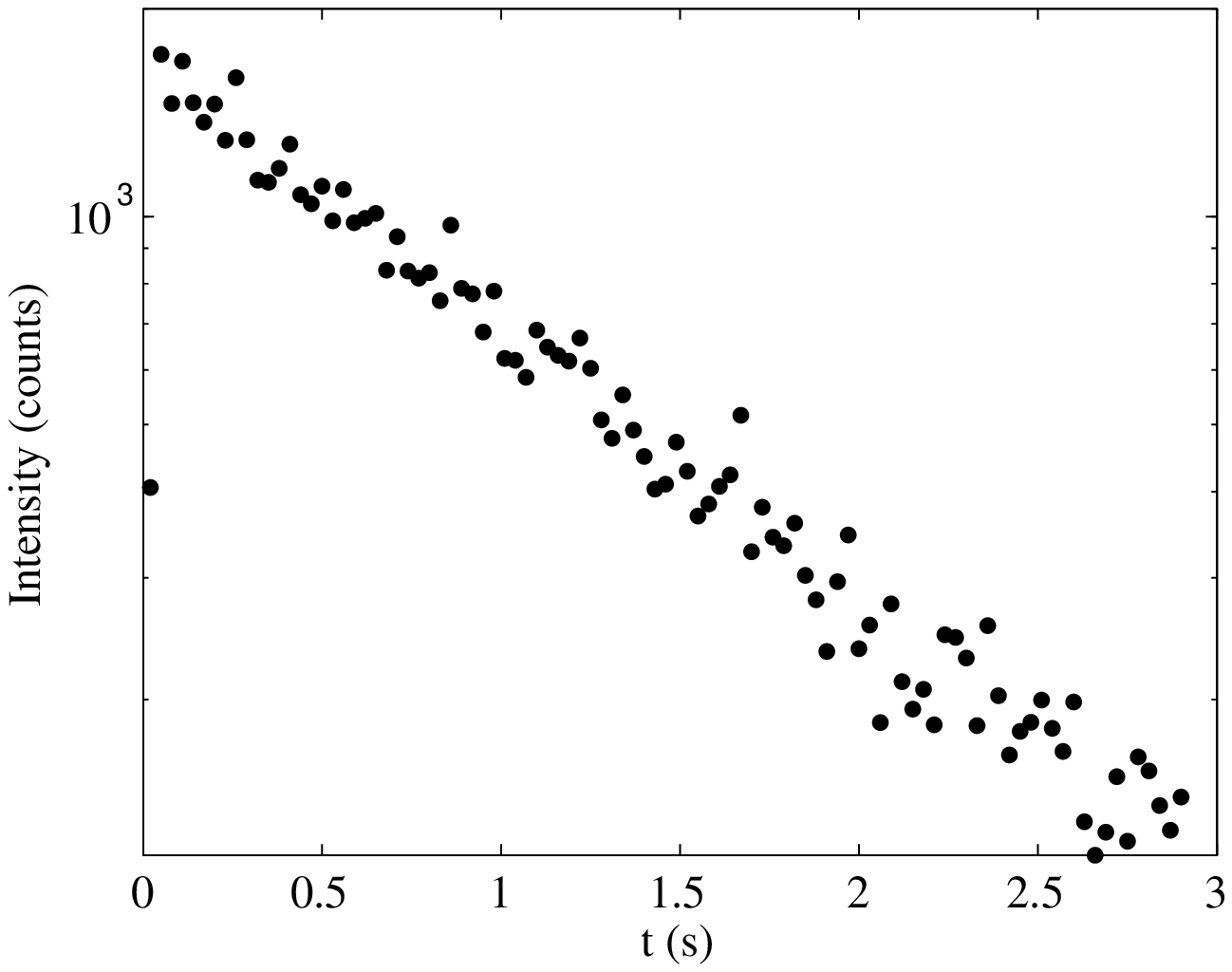}}  
\caption{Decay curve for the a\ls[2]{3}{P} level at base pressure.  \label{fig:decaycurve}}
\resizebox{\columnwidth}{!}{\includegraphics{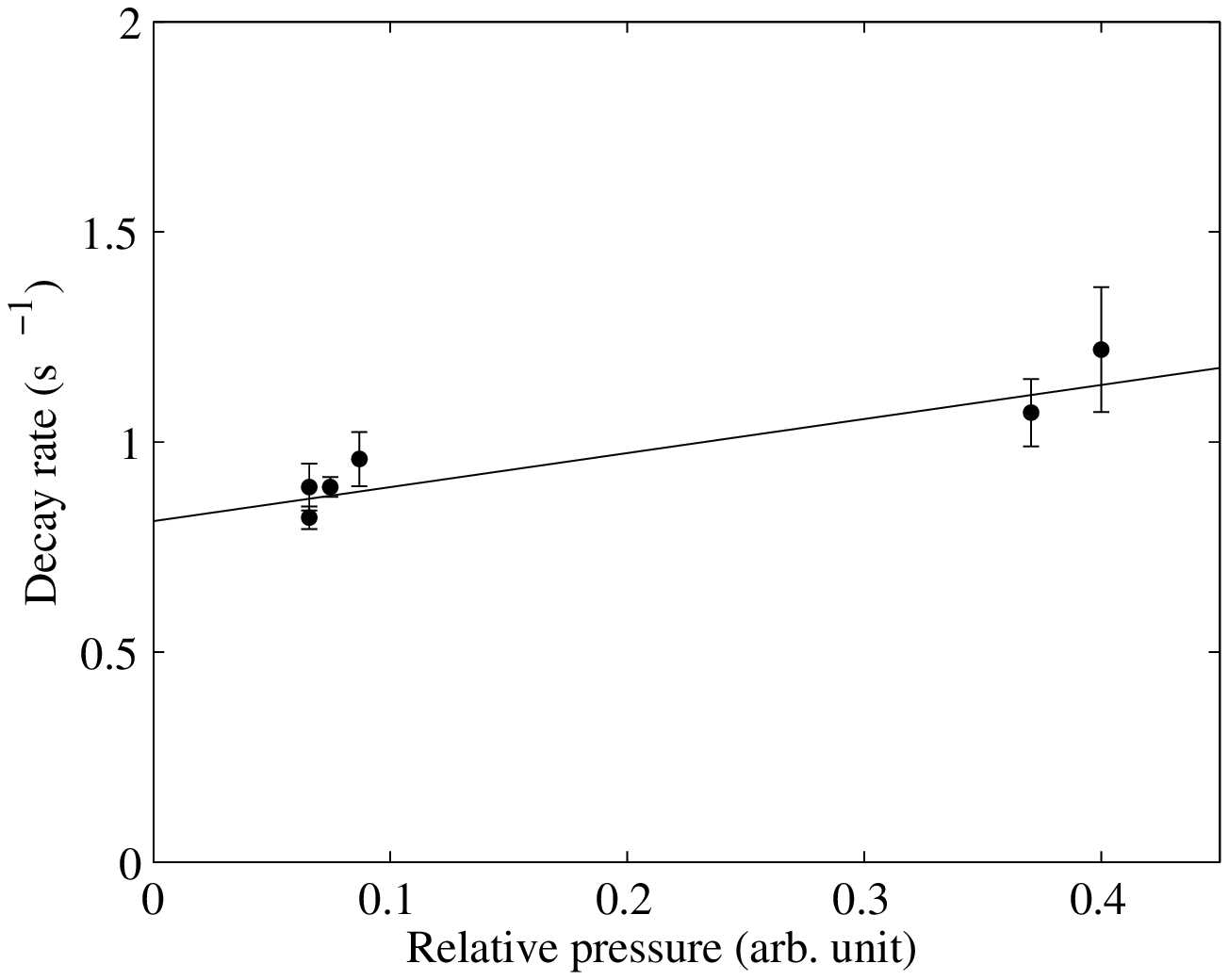}}  
\caption{Stern-Vollmer plot for the a\ls[2]{3}{P} level. The apparent lifetime is measured at base pressure and raised pressure, and the true radiative lifetime is derived by extrapolation to zero pressure. \label{fig:sv}}
\end{figure}

The Sc$^+$ ions are produced in an ion source and, after being accelerated to an energy of 40 keV and passing through an isotope-selecting bending magnet, are injected into the storage ring. A continuous wave (cw) tunable dye laser, with a mechanical chopper, is used to probe the stored ions. The wavelength is tuned to match a strong transition from the metastable level to an upper level with opposite parity (Fig.~\ref{fig:pump}). The subsequent decay is fast, a few nanoseconds. The intensity of the prompt decay, the fluorescence, is proportional to the number of ions in the metastable level probed at a certain delay after injection into the ring. By varying the delay time of the laser pulse after injection into the ring, the time dependence of the population in the metastable level is monitored. For each injection, the population at one time is measured and corresponds to one point on the decay curve. Subsequent injections into the ring are probed at different time delays, and in this way the decay curve is built up. The technique is destructive in the sense that once a laser pulse is applied, the level is emptied and a new set of ions must to be injected for the next measurement. It is thus important that the number of ions in the level measured do not vary too much and that the variations can be monitored. During the measurement, a number of properties, such as the beam lifetime, number of ions in each injection, and the fraction of ions in the level of interest, are monitored to allow for corrections of possible drifts and changes in the ion production \citep{M03}. A corrected decay curve is shown in Fig.~\ref{fig:decaycurve}.

Despite the ultra-high vacuum, the stored ions are affected by collisions with rest gas. Both collisional deexcitation (quenching) and excitation (repopulation) are observed and corrected for. A detailed description of the corrections is given in previous publications (see above). The quenching is observed as a pressure dependence in the measured lifetime. The collisions add a decay rate in addition to the radiative rate of interest. This collisional decay rate can be subtracted by measuring the lifetime as a function of pressure and extrapolating to zero pressure in a Stern-Vollmer plot \citep{D96a}, see Fig.~\ref{fig:sv}.

The experimental lifetimes of 3d$^2$~a\ls[0,1,2]{3}{P} are presented in Table 1 along with uncertainties, which are one sigma uncertainties to the linear fit of the Stern-Vollmer plot.

\begin{table}
\caption{Lifetimes in seconds.\label{tab:lifetime}}
\label{tab:lifetimes}
\begin{tabular}{lcccc} \hline
 & & \multicolumn{2}{c}{This work} & WK69$^a$ \\
   & & Experimental$^b$    & Theory & \\ \cline{1-2} \cline{3-5}
a\ls[0]{3}{P} & & 1.28$\pm0.13$  & 1.44   & 1.07     \\
a\ls[1]{3}{P} & & 1.42$\pm0.22$ & 1.43   & 1.07     \\ 
a\ls[2]{3}{P} & & 1.24$\pm0.04$  & 1.43   & 1.07     \\ \hline
\end{tabular}
\\$^a$ \citet{WK69}
\\$^b$ The uncertainty quoted is a one standard deviation of the linear fit to the Stern-Vollmer plot.
\end{table}

\begin{figure}
\resizebox{\columnwidth}{!}{\includegraphics{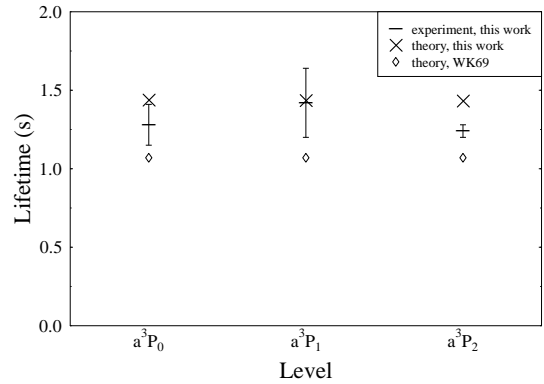}}  
\caption{Comparisons of experimental and theoretical lifetimes for the 3d$^2$~a$^3$P levels from the present work, and to previous calculations by \citet{WK69}. \label{fig:tauplot}}
\end{figure}

\begin{figure}
\resizebox{\columnwidth}{!}{\includegraphics{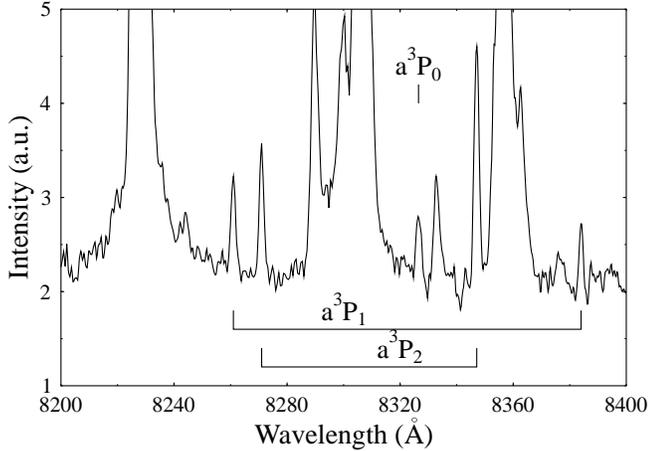}}  
\caption{Spectrum showing the [\ion{Sc}{ii}] lines used in the present analysis. The upper level of each group is indicated. The transitions are listed in Table \ref{tab:aresult}.\label{fig:scspec}}
\end{figure}

\subsection{Branching fractions for \em{3d4s~a\ls{3}{D}-3d$^2$~a\ls{3}{P}}}
With the high spatial resolution of the STIS spectrograph onboard $HST$, the different parts of the nebulosity and the ejecta around Eta Carinae can be spatially resolved. Many of these have low densities, and allow for parity forbidden lines to be produced. One region rich in forbidden lines is the so-called Strontium Filament. 

Among the many forbidden iron-group elements lines, the a\ls{3}{D}-a\ls{3}{P} transitions in [\ion{Sc}{ii}] are observed, as shown in Fig.~\ref{fig:scspec}. Other forbidden lines of [\ion{Sc}{ii}] fall longward of the observed wavelength region. Lines from a\ls[4]{1}{G}, with energy similar to the a\ls{3}{P} levels, are masked in the spectrum by other strong lines. The major decay from b\ls[0]{1}{S} would have a wavelength around 4270~\AA, but this line is not observed, probably due to high excitation energy and thus low population of the b\ls[0]{1}{S} level. A problem that occurs when using astrophysical spectra instead of laboratory sources to measure BFs is the blending of lines from other elements. The present lines fall in the region where the HI Paschen series has numerous lines, but from other spatial regions of the nebula. However, theser are weak and can be avoided. The 8263\AA\ [\ion{Sc}{ii}] line is corrected for blending with an [\ion{Fe}{ii}] line, which contributes to the observed emission feature. The amount of blending is determined to be 10\% of the total intensity, and is derived using other [\ion{Fe}{ii}] lines, that are strong in other spatial regions along the line-of-sight. The a\ls[1,2]{3}{P} each have two major transitions each, whereas a\ls[0]{3}{P} has one single dominant decay channel. Astrophysical spectra need to be corrected for interstellar reddening, but since the lines span a very limited wavelength region, 8250-8380\AA, this is not necessary for the present lines.

\section{Calculations} \label{theory}

We have undertaken an extensive calculation of rates of forbidden transitions
in \ion{Sc}{ii} using configuration interaction wave functions, which are expressed
in the form
\begin{equation}
\Psi({LSJ}) = \sum_{i=1}^{M} a_i \Phi_i(\alpha_i{LSJ})
\label{eqn:wf}
\end{equation}
where  $\{\Phi_i\}$ are single-configuration functions (CSFs) and the expansions,
in general, include summations over $L$ and $S$, thus allowing mixing between
different $LS$ terms with the same $J$ value.  For a specific choice of
$\{\Phi_i\}$, the expansion coefficients $\{a_i\}$ are the eigenvector
components of the Hamiltonian matrix elements $H_{ij}$ = $<\Phi_i|H|\Phi_j>$.  
The relativistic effects that make these forbidden transitions possible were
incorporated into the calculation using the Breit-Pauli approximation, which
should be adequate for moderately heavy ions such as \ion{Sc}{ii}.  Specifically,
we included in our Hamiltonian the one-body mass-correction and Darwin terms
and the nuclear spin-orbit interaction, as well as the two-body spin-other-orbit
and spin-spin terms.

The states involved in this calculation are normally labelled 3d$^2$, 3d4s, or 4s$^2$.
However, the optimal functions for these states differ quite substantially, so, for example,
the optimal 3d orbital for 3d$^2$ states is quite different from that for 3d4s states.
Our calculations assume that we are using a common set of orthogonal orbitals to describe
all the states and we need to allow for this effective non-orthogonality by introducing
additional, correcting orbitals. 

The calculations described here have been undertaken with the code CIV3 
\citep{H75,GH78,HGF91}. 
For the 1s,2s,2p,3s,3p, and 4s orbitals we used the Hartree-Fock radial functions given
by Clementi and Roetti (1974) for the [1s$^2$2s$^2$2p$^6$3s$^2$3p$^6$]3d4s $^3$D state.
We optimised 3d on the 3d$^2$ $^3$P state rather than 3d4s $^3$D to provide
a balance between different states and because the correcting orbital 4d can be then
treated as a linear correction rather than quadratic.  Finally, 4p,4f, and 5d are correlation
orbitals for the outer subshell.  
The full process of optimising the radial functions and 
descriptions of the configuration sets will be given in a separate paper (Lundin et al., in prep)

The full set of configurations used, for all $LS$ symmetries, is shown in Table~\ref{tab:conf_used}.
They comprise all the configurations that can be constructed for valence shell correlation,
together with the important 3p$^2 \rightarrow $3d$^2$ correlation effect in the dominant
configurations (3d$^2$ and 3d4s, together with the effects of the correction orbitals
that give rise to 3d4d and 3d5d).  We have also allowed the main effect of the replacement 
of 3p$^2$ by two other orbital functions.  The even parity $LS$ symmetries in our work are $^{3,1}$D, 
$^3$P, $^1$S, $^3$F and $^1$G, appropriately mixed for each $J$-value.  In this way,
we are able to determine wave functions for each of the 14 levels normally labelled with 3d$^2$, 3d4s,
or 4s$^2$.  

To improve the calculations, we make small corrections to the diagonal elements
of the Hamiltonian matrix \citep{H96} so that the final calculated energy separations agree with the experimental 
values \citep{JL80}.  We have termed this process `fine-tuning'.  We have found that it improves 
the accuracy of the mixing coefficients ($a_i$~in~(\ref{eqn:wf})) 
and hence the calculated transition rates.  We find that there is
very little mixing between states of different $LS$, especially for the triplets studied in this paper. 

\begin{table}
\caption{Configurations used in full calculation  \label{tab:conf_used}}
\begin{tabular}{ll@{\extracolsep{1.0cm}}l@{\extracolsep{0.4cm}}l} 
\hline \\[-0.2cm]
3p$^6$ 3d$^2$ & 3p$^6$ 4p$^2$ & 3p$^4$ 3d$^4$     & 3p$^4$ 3d$^2$4s5d \\ 
3p$^6$ 3d4s & 3p$^6$ 4p4f & 3p$^4$ 3d$^3$4s   & 3p$^4$ 3d$^2$4p4f \\
3p$^6$ 3d4d & 3p$^6$ 4d$^2$ & 3p$^4$ 3d$^3$4d   & 3p$^4$ 3d$^2$4p$^2$ \\
3p$^6$ 3d5d & 3p$^6$ 4d5d & 3p$^4$ 3d$^3$5d   & 3p$^4$ 3d$^2$4d$^2$ \\
3p$^6$ 4s$^2$ & 3p$^6$ 4f$^2$ & 3p$^4$ 3d$^2$4s$^2$ & 3p$^4$ 3d$^2$4f$^2$ \\
3p$^6$ 4s4d & 3p$^6$ 5d$^2$ & 3p$^4$ 3d$^2$4s4d & 3p$^4$ 3d$^2$5d$^2$ \\
3p$^6$ 4s5d &&& \\
\hline
\end{tabular}
\end{table}

\begin{table*}
\caption{The strongest lines from the a\ls{3}{P} levels. The lines lacking experimental branching fractions (BFs) are either too weak to be detected or have wavelengths outside the observed spectral region. The residual is the total contribution from the transitions from the upper level not included in the table. \label{tab:aresult}}
\begin{tabular}{ccrcclll} \hline \hline
upper level     & lower level     & $\lambda_{vac}$ (\AA) & BF$_{exp}$ (\%) & BF$_{th}$ (\%) & A$_{\mathrm{exp}}$ (s$^{-1}$)$^a$ & A$_{\mathrm{theory}}$ (s$^{-1}$) \\ \hline
a\ls[0]{3}{P}	& a$^3$F$_2$	  & 13752.83       &                 & 4.34           &                                   & 0.0302                           \\
		& a$^3$D$_2$	  & 8328.91 	   & 95.6$\pm$4     & 95.6           & 0.75$\pm0.08$                           & 0.666                            \\ 
		& \multicolumn{2}{l}{Weaker transitions (total)}                      & 0.06           &                                   &                 \\[3mm]
a\ls[1]{3}{P}   & a$^3$F$_2$	  & 13701.20       &                 & 1.48           &                                   & 0.0103                           \\
		& a$^3$F$_3$	  & 13854.39       &                 & 2.79           &                                   & 0.0195                           \\
		& a$^3$D$_2$	  & 8309.94        &                 & 7.98           &                                   & 0.0557                           \\
		& a$^3$D$_3$	  & 8386.63        & 40.0$\pm$6     & 43.0           & 0.28$\pm0.06$                           & 0.300                            \\ 
		& a$^3$D$_1$	  & 8263.44        & 47.7$\pm$7     & 44.7           & 0.34$\pm0.07$                           & 0.312                            \\ 
		& \multicolumn{2}{l}{Weaker transitions (total)}                      & 0.05           &                                   &                 \\[3mm]
a\ls[2]{3}{P}   & a$^3$F$_3$	  & 13753.55       &                 & 0.87           &                                   & 0.0061                           \\
		& a$^3$F$_4$	  & 13953.56       &                 & 3.10           &                                   & 0.0217                           \\
		& a$^3$D$_2$	  & 8273.56        & 27.8$\pm$5     & 34.2           & 0.23$\pm0.04$                           & 0.239                            \\
		& a$^3$D$_3$	  & 8349.57        & 58.7$\pm$8     & 52.3           & 0.47$\pm0.07$                           & 0.366                            \\
		& a$^3$D$_1$	  & 8227.46        &                 & 9.09           &                                   & 0.0636                           \\
		& \multicolumn{2}{l}{Weaker transitions (total)}                      & 0.04           &                                   &                 \\
\hline                                
\end{tabular}                 
\\$^a$The uncertainty is derived from the uncertainty of the lifetimes and the branching fractions.
\end{table*}

\begin{figure}
\resizebox{\columnwidth}{!}{\includegraphics{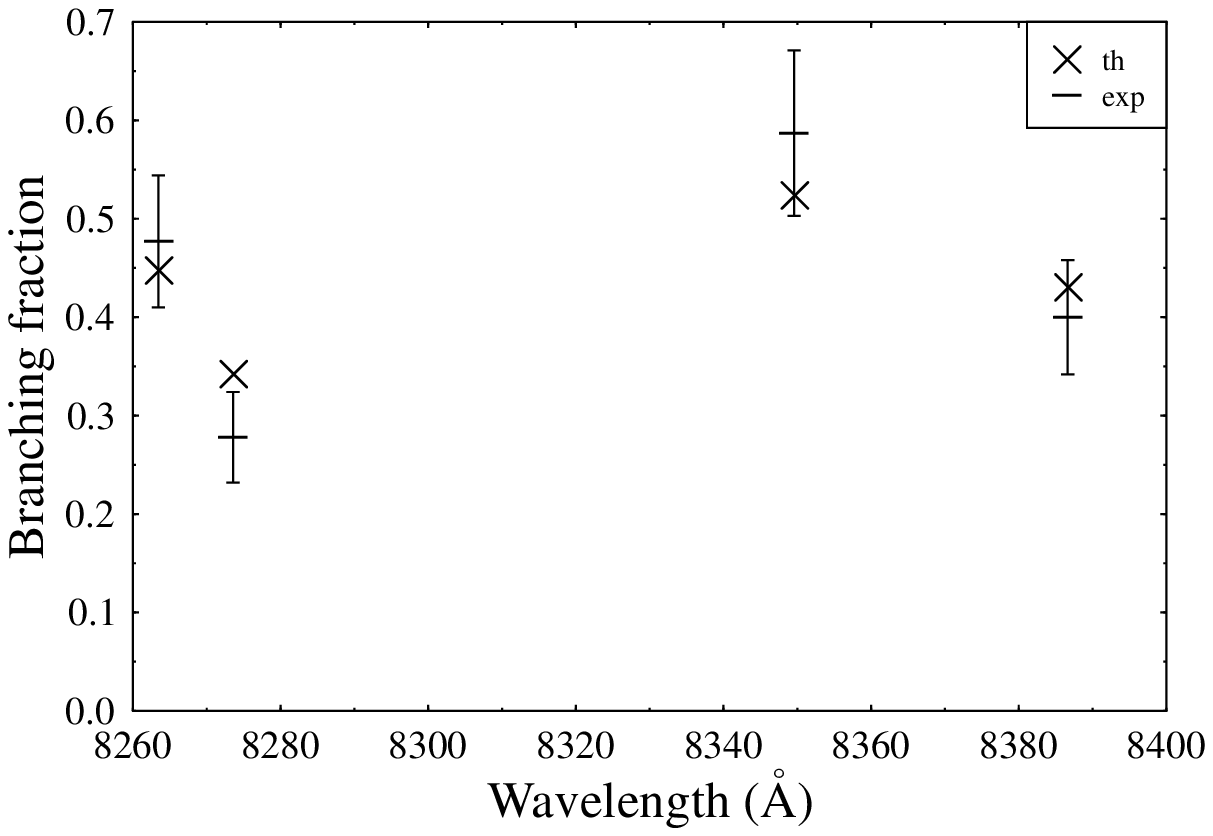}}  
\caption{Branching fractions (BF) for 3d4s\,a$^3$D~-~3d$^2$\,a$^3$P. The experimental BFs are derived from the nebular spectrum, using the theoretical values for the residual lines. The 8328\,\AA\ line from a$^3$P$_0$ is not included since this is the only decay observed from this level, and thus makes the experimental BF uninteresting. \label{fig:bfresult}}
\resizebox{\columnwidth}{!}{\includegraphics{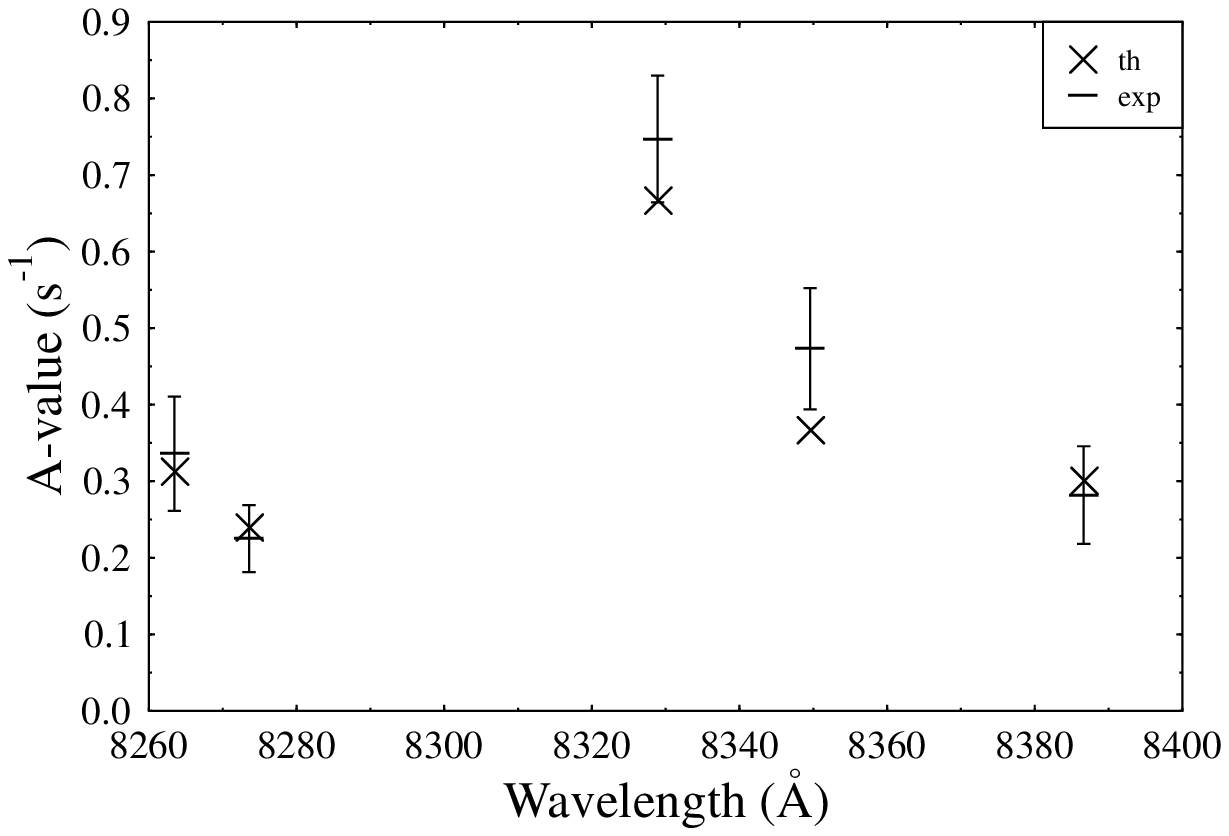}}   
\caption{Transition rates ($A$-values) for the 3d4s~a$^3$D-3d$^2$~a$^3$P transitions. The difference between the theoretical and expermental values, compared to Fig.~4, is due to the difference in lifetimes between experiment and calculations.\label{fig:aresult}}
\end{figure}

\section{Results} \label{results}
The measured and calculated lifetimes are given and compared in Table~\ref{tab:lifetimes} and Fig.~\ref{fig:tauplot}. Our experimental values are 1.28, 1.42, and 1.24~s for the 3d$^2$~a\ls{3}{P} levels, for $J$=0,1,2, respectively, and our calculated values are 1.44, 1.43, and 1.43~s for the same levels. E2 transitions dominate the calculation of the lifetimes of the 3d$^2$ levels. The previous calculation by \citet{WK69} gives 1.07~s for all levels. The Warner and Kirkpatrick calculation is based on a single-configuration approximation.  This neglects configuration interaction, which for \ion{Sc}{ii} can be important.  The energy levels are obtained semi-empirically (effectively experimental transition energies are used) and the mixing between different $LS$ terms is therefore approximately correct. The radial integrals of the dipole matrix elements are obtained using scaled Thomas-Fermi orbitals, and are therefore obtained without the use of the variational principle.  Hence their results, while being of reasonable accuracy, are not so reliable as the present calculations.

The uncertainty for the experimental lifetimes are 4\%-15\%, and presented in Table~1. The uncertainty quoted is one standard deviation of the linear fit to the Stern-Vollmer plot. The variations in uncertainty between the three investigated levels are mainly due to variations in experimental conditions. We measured the lifetime of a\ls[2]{3}{P} separately from the lifetimes of the a\ls[0,1]{3}{P} levels. During the study of the a\ls[2]{3}{P} level, more favourable experimental conditions were achieved. This included a higher ion current in the storage ring as well as higher power from the ring dye laser which in turn resulted in better statistics with smaller uncertainties.

The smaller uncertainty associated with the lifetime of the $^3$P$_2$ level is also a consequence of the fact that the we measured the lifetime of this level in a wider relative pressure range compared to the measurements of the $^3$P$_{0,1}$ levels, which makes the extrapolation to zero pressure more certain. It should, however, be emphasized that in all three measurements, the effect of collisional quenching is small since the differences between the deduced pure radiative lifetimes and the longest individual lifetimes measured at base pressure is 2-11\% .

The measured branching fractions are presented with error bars in Fig.~\ref{fig:bfresult}, along with the theoretical values. The a\ls[0]{3}{P} value is not included since only one line carries the major intensity. This means that no experimental BF can be determined for lines from this level.
The uncertainties of the BFs are in the range 4-16\% and include contributions from intensity measurement, line deblending and correction for unobserved lines (residual). The residual is the total BF for all unmeasureable lines:\\ 
\hspace*{5mm}i) either falling outside the observed wavelength region, or \\
\hspace*{5mm}ii) too weak to be measured.\\
The residuals for a\ls[0,1,2]{3}{P} are, including the transitions to a\ls{3}{F},  0.05, 0.12, 0.14, respectively, and an uncertainty of 100\% is assigned for these values.  The agreement with the calculated values is within the 1-sigma error bars for most transitions. 

The resulting experimental $A$-values, derived from the measured lifetimes and astrophysical BFs, are presented in Table \ref{tab:aresult} and Fig.~\ref{fig:aresult} and are in the range 0.23 to 0.75 s$^{-1}$. The uncertainties of the experimental transition rates are in the range 11-21\%, and result from uncertainties in the lifetimes and in the BFs, including uncertainty of unmeasureable lines (the residual, discussed above). The contributions to the total uncertainty from the lifetime and branching fractions are of similar importance.
There is relatively little configuration mixing between the states, so we would expect the level of accuracy of the calculated values to be comparable with that encountered in E1 transitions for which there is little mixing. For the theoretical values, an accuracy of 5-10\% could reasonably be assigned.

\section{Conclusions}
We have measured lifetimes of the 3d$^2$ $^3$P$_{0,1,2}$ levels in \ion{Sc}{ii}. By combining these with astrophysical branching fractions from $HST$/STIS spectra form Eta Carinae, experimental transition rates are derived for the parity forbidden [\ion{Sc}{ii}] lines 3d4s\,a$^3$D~-~3d$^2$\,a$^3$P. The values are compared with new CIV3 calculations, and the agreement is good.

\acknowledgements
This work is supported by the Swedish Research Council (Vetenskapsrådet) and is part of a project funded by the Swedish National Space Board (SNSB). We are grateful to the people at the CRYRING facility at MSL (Stockholm University) for providing an excellent laboratory environment. One of us (AH) is pleased to acknowledge support from PPARC (UK) through the award of Rolling Grant PP/D00103X/1.The data of Eta Carinae used in this paper were obtained through the $HST$ program 8619 (P.I.\ Kris Davidson), and we are grateful to Dr Theodore Gull for providing calibrated spectra.
This research has made use of NASA's Astrophysics Data System Bibliographic Services.
\bibliographystyle{aa}
\bibliography{d:/latex/bibtex/hartman} 

\begin{thebibliography}{19}
\expandafter\ifx\csname natexlab\endcsname\relax\def\natexlab#1{#1}\fi

\bibitem[{{Bautista} {et~al.}(2002){Bautista}, {Gull}, {Ishibashi}, {Hartman},
  \& {Davidson}}]{BGI02}
{Bautista}, M.~A., {Gull}, T.~R., {Ishibashi}, K., {Hartman}, H., \&
  {Davidson}, K. 2002, MNRAS, 331, 875

\bibitem[{{Bautista} {et~al.}(2006){Bautista}, {Hartman}, {Gull}, {Smith}, \&
  {Lodders}}]{BHG06}
{Bautista}, M.~A., {Hartman}, H., {Gull}, T.~R., {Smith}, N., \& {Lodders}, K.
  2006, \mnras, 370, 1991

\bibitem[{{Davidson} \& {Humphreys}(1997)}]{DH97}
{Davidson}, K. \& {Humphreys}, R.~M. 1997, ARA\&A, 35, 1

\bibitem[{{Demtr\"oder}(1996)}]{D96a}
{Demtr\"oder}, W. 1996, {Laser spectroscopy. Basic concepts and
  instrumentation, 2nd ed} (New York ; Berlin : Springer, cop. 1996)

\bibitem[{{Glass} \& {Hibbert}(1978)}]{GH78}
{Glass}, R. \& {Hibbert}, A. 1978, Comp.\ Phys.\ Commun.\, 16, 19

\bibitem[{{Hartman} {et~al.}(2004){Hartman}, {Gull}, {Johansson}, {Smith}, \&
  {HST Eta Carinae Treasury Project Team}}]{HGJ04}
{Hartman}, H., {Gull}, T., {Johansson}, S., {Smith}, N., \& {HST Eta Carinae
  Treasury Project Team}. 2004, A\&A, 419, 215

\bibitem[{{Hartman} {et~al.}(2005){Hartman}, {Schef}, {Lundin}, {Ellmann},
  {Johansson}, {Lundberg}, {Mannervik}, {Norlin}, {Rostohar}, \&
  {Royen}}]{HSL05}
{Hartman}, H., {Schef}, P., {Lundin}, P., {et~al.} 2005, \mnras, 361, 206

\bibitem[{Hibbert(1975)}]{H75}
Hibbert, A. 1975, Comput. Phys. Commun., 9, 141

\bibitem[{Hibbert(1996)}]{H96}
Hibbert, A. 1996, Physica Scripta, T65, 104

\bibitem[{Hibbert {et~al.}(1991)Hibbert, Glass, \& Froese~Fischer}]{HGF91}
Hibbert, A., Glass, R., \& Froese~Fischer, C. 1991, Comput. Phys. Commun., 64,
  455

\bibitem[{{Johansson} {et~al.}(2002){Johansson}, {Derkatch}, {Donnelly},
  {Hartman}, {Hibbert}, {Karlsson}, {Kock}, {Li}, {Leckrone}, \& {Litz{\'
  e}n}}]{JDD02}
{Johansson}, S., {Derkatch}, A., {Donnelly}, M.~P., {et~al.} 2002, Physica
  Scripta, T100, 71

\bibitem[{{Johansson} \& {Litz{\'e}n}(1980)}]{JL80}
{Johansson}, S. \& {Litz{\'e}n}, U. 1980, \physscr, 22, 49

\bibitem[{{Lidberg} {et~al.}(1999){Lidberg}, {Al-Khalili}, {Norlin}, {Royen},
  {Tordoir}, \& {Mannervik}}]{LAN99}
{Lidberg}, J., {Al-Khalili}, A., {Norlin}, L.-O., {et~al.} 1999, Nuclear
  Instruments and Methods in Physics Research B, 152, 157

\bibitem[{{Mannervik}(2003)}]{M03}
{Mannervik}, S. 2003, Physica Scripta Volume T, 105, 67

\bibitem[{{Mannervik} {et~al.}(2005){Mannervik}, {Ellmann}, {Lundin}, {Norlin},
  {Rostohar}, {Royen}, \& {Schef}}]{MEL05}
{Mannervik}, S., {Ellmann}, A., {Lundin}, P., {et~al.} 2005, Physica Scripta
  Volume T, 119, 49

\bibitem[{{Moore}(1945)}]{M45}
{Moore}, C.~E. 1945, {A multiplet table of astrophysical interest.} (Princeton,
  N.J., The Observatory, 1945.~Rev.~ed.)

\bibitem[{{Smith} {et~al.}(2004){Smith}, {Morse}, {Gull}, {Hillier}, {Gehrz},
  {Walborn}, {Bautista}, {Collins}, {Corcoran}, {Damineli}, {Hamann},
  {Hartman}, {Johansson}, {Stahl}, \& {Weis}}]{SMG04}
{Smith}, N., {Morse}, J.~A., {Gull}, T.~R., {et~al.} 2004, ApJ, 605, 405

\bibitem[{{Warner} \& {Kirkpatrick}(1969)}]{WK69}
{Warner}, B. \& {Kirkpatrick}, R.~C. 1969, \mnras, 144, 397

\bibitem[{{Zethson} {et~al.}(2001){Zethson}, {Gull}, {Hartman}, {Johansson},
  {Davidson}, \& {Ishibashi}}]{ZGH01}
{Zethson}, T., {Gull}, T.~R., {Hartman}, H., {et~al.} 2001, AJ, 122, 322

\end{thebibliography}
\end{document}